\documentclass[12pt]{iopart}
\usepackage{epsfig}
\newcommand{\bea}{ \begin{eqnarray} }
\newcommand{\eea}{ \end{eqnarray}  }
\newcommand{\bc}{ \begin{center}  }
\newcommand{\ec}{ \end{center}  }

\newcommand{\vks}{\mbox{$V_{\mbox{\scriptsize{KS}}}(\vec{r})$}}
\newcommand{\vjel}{\mbox{$V_{\mbox{\scriptsize{jel}}}(\vec{r})$}}
\newcommand{\vd}{\mbox{$V_{\mbox{\scriptsize{d}}}[\rho(\vec{r})]$}}
\newcommand{\vxc}{\mbox{$V_{\mbox{\scriptsize{xc}}}[\rho(\vec{r})]$}}
\newcommand{\vxci}{\mbox{$V_{\mbox{\scriptsize{xc}}}[\rho_i(\vec{r})]$}}
\newcommand{\vxcts}{\mbox{$V_{\mbox{\scriptsize{xc}}}[\rho_{2s}(\vec{r})]$}}
\newcommand{\vsic}{\mbox{$V_{\mbox{\scriptsize{SIC}}}^i(\vec{r})$}}
\begin{document}

\letter{Spurious oscillations from local self-interaction correction  
in high energy photoionization calculations for metal clusters
}

\author{M.E. Madjet, Himadri S. Chakraborty  
\footnote[3]{To
whom correspondence should be addressed (himadri@mpipks-dresden.mpg.de)}
, and Jan-M. Rost
}

\address{Max-Planck-Institut f\"{u}r Physik Komplexer Systeme,  
N\"{o}thnitzer Strasse 38, D-01187 Dresden, Germany}

\begin{abstract}
We find that for simple metal clusters a single-electron description of the ground state 
employing self-interaction correction (SIC) in the framework of local-density
approximation strongly contaminates  
the high energy photoionization cross sections with spurious oscillations 
for a subshell containing node(s). This effect is shown connected
to the unphysical structure that SIC generates 
in ensuing state-dependent radial potentials around a position
where the respective orbital density attains nodal zero. 
Non-local Hartree-Fock that {\em exactly} eliminates the electron self-interaction
is found entirely free from this effect. 
It is inferred that while SIC is largely unimportant in high photon-energies, 
any implementation of it within the local frame 
can induce unphysical oscillations in the high energy photospectra
of metal clusters pointing to a general need for caution in choosing appropriate theoretical
tools.
\end{abstract}

%Density Functional Theory
%Electronic and magnetic properties of clusters
%Optical properties of clusters
\pacs{31.15.Ew, 36.40.Cg, 36.40.Vz}

%Journal of Physics B

%\maketitle
\noindent
The local-density approximation (LDA), along with its time-dependent version, 
is a standard theoretical 
technique to describe the structure and dynamics of large systems. From a practical
standpoint, LDA is typically 
preferred to other conventional many-body methods (such as, Hartree-Fock (HF)
or techniques based on configuration-interactions) because of its relatively low 
computational costs.  
In the context of the studies involving static and dynamical properties of
simple metal clusters LDA has proved to be particularly successful [1-3].
However, a well known
drawback of LDA is that it 
only partially accounts for unphysical electron self-interactions.
As a consequence, the resulting potential for a finite system 
decays exponentially at large distance instead of producing the physical $1/r$ behavior.
To render the long distance behavior of the LDA potential realistic, therefore, 
approximation schemes have been suggested [4]. The most general and widely
applied to remedy the error is the one proposed by Perdew and Zunger [5],
which concerns an orbit-by-orbit
elimination of self-interaction, although the scheme
immediately makes the potential state-dependent.
The self-interaction corrected LDA (LDA-SIC)   
improves remarkably the vast variety of results related to many structural properties 
of physical systems: for instance, improvements in total energies of atoms, 
allowance for self-consistent bound solutions for negative ions, prediction
of orbital energies that are close to electron removal energies thus restoring
Koopmans' theorem, ensuring dissociation of heteronuclear molecules to
neutral fragments, improvement of the band gap in solids etc.\ 
(a good account in this regard may be found through Ref.\ 4). 
In the dynamical regime too, especially in the context of low-energy
photoionization of simple metal clusters, the description of the electronic ground state 
via LDA-SIC results in important many-body effects including single electron Rydberg
resonances [6,7].
 
At photon-energies well beyond the ionization threshold 
the photospectrum shows special qualitative behavior.
For spherical jellium clusters  
over this energy range theory predicts a characteristic oscillation in the 
cross section with a frequency connected to the cluster diameter. 
The mechanism behind this oscillatory pattern is  
the interference of electron waves emanated from equivalent sites of the cluster edge [8]. 
While there has been no experimental study on metal clusters, 
oscillations in the photoelectron intensity are indeed observed 
for fullerene molecules [9].
Generically, the high energy photoionization process should be
rather sensitive to the degree of
accuracy in the description of the ground state. This can
be understood from the fact that in an independent particle model the high-energy
transition matrix element has a leading contribution from the Fourier
transform of the ground state wavefunction to the photoelectron momentum space (or
retarded-momentum space if non-dipole interactions are included) [10]. 
From such an elementary viewpoint, LDA-SIC may also appear to be a suitable 
tool for the high energy 
photoionization studies of various cluster systems.
However, this paper shows that while the correction for self-interaction
is virtually unimportant in the study of energetic photoionization of metal
clusters, any approximation to it in a local frame can generate spurious oscillations
in the cross section for photoelectrons emerging from subshells having orbital node(s).    
The point is illustrated by presenting calculations  
on Na$_{20}$, which can be well described by a spherical
jellium model, and which is the smallest system ($1s^21p^61d^{10}2s^2$)
having 
one subshell ($2s$) with a node. 

The usual single electron potential in the Kohn-Sham LDA formalism is 
\bea
\vks &=& \vjel+\vd+\vxc
\eea
where the terms on the right-hand-side are respectively jellium, 
direct (Hartree) $\vd
= \int d\vec{r}' \rho(\vec{r}')/|\vec{r}-\vec{r}'|$, and
exchange-correlation potentials. The ground state 
electronic density $\rho (\vec{r})$ is defined in terms of
single-electron densities $\rho_i$ and orbitals $\phi_i$:
$$
\rho (\vec{r}) = \sum_{i=1}^N \rho_i (\vec{r}) = \sum_i |\phi_i (\vec{r})|^2 
$$
As mentioned earlier, an approximate prescription for SIC to this LDA potential (1) is to  
eliminate orbitalwise from the outset those terms which represent an electron $i$
interacting to itself. The resulting 
orbital-specific potentials, therefore, are 
\bea
\vsic &=& \vjel + \int d\vec{r}'\frac{\rho (\vec{r}')-\rho_i (\vec{r}')}{|\vec{r}-\vec{r}'|}
         + \vxc -\vxci
\eea 
As the exact form of $V_{\mbox{\small xc}}$ is unknown a widely used scheme
is to employ the formula[11]:
\bea
\vxc &=& -\left(\frac{3\rho(\vec{r})}{\pi}\right)^{1/3} - 
        0.0333 \log \left[1+11.4\left(\frac{4\pi\rho(\vec{r})}{3}\right)^{1/3}\right]
\eea
The first term on the right-hand-side 
in the above expression is exactly
derivable by a variational approach from  
the HF exchange energy of a uniform electron system with a uniform positively charged background;
the second term is the so called correlation potential, a quantity
not borne in HF formalism.  
We use LDA potentials both with and without SIC approximation to calculate
the dipole photoionization cross sections  upto approximately 1 KeV photon-energy for 
each subshell of the Na$_{20}$ cluster in the
independent particle frame\footnote[1]{Of course at such high energy  
the Na$^+$ core will ionize. However, the inclusion of this effect, 
going beyond the jellium frame, will not change our result 
qualitatively.}. Calculations are also performed in 
the self-consistent HF scheme to better identify the origin of the resulting
discrepancy between the two LDA predictions.  
Quantities
are in atomic units throughout, except where specified otherwise.

LDA and LDA-SIC cross sections for each of the $1s$, $1p$ and $1d$ subshells are found to be 
almost identical at high enough energies showing a single monotonic 
oscillation. Results using HF for these subshells 
yield similar qualitative behavior. The situation, however, is quite different 
for the $2s$ photoionization. Figure 1 presents $2s$ cross sections as obtained 
through LDA, LDA-SIC, and   
HF, as a function of $2s$ photoelectron momentum $k_{2s} = \sqrt{2(E-I_{2s})}$,
with $I_{2s}$ ($\sim$ 3.5 eV) being the $2s$ ionization threshold. 
Generally, 
in the low-energy range for all subshells of Na$_{20}$
HF predictions are different from LDA owing to the partly
non-identical ground state correlation
they account for and this causes a constant phase difference
between them at higher energies, where such correlation effects are
insignificant.
Bearing this in mind we find in figure 1 that
while LDA and HF again maintain the same trend oscillationwise, LDA-SIC 
points to a progressively strong qualitative difference starting roughly
from 40 eV photon-energy. 
To identify closely the
discrepancy between $\sigma_{2s}$ with and without SIC we have evaluated the 
Fourier transforms of $\sigma_{2s}
(k_{2s})$ (see figure 2). 
Both LDA and HF are seen to have approximately the same Fourier spectrum 
with just one peak. But LDA-SIC contains three additional peaks beside 
the one that is common to all three spectra. This common frequency  
is connected to the diameter of 
the cluster. In fact, a simple theoretical analysis shows that high energy 
photo cross sections of a spherical jellium cluster oscillate in the respective
photoelectron momentum space at a frequency $2R_{\mbox{c}}$, where $R_{\mbox{c}}$    
is the cluster radius [8]. However, where do the other frequencies in the LDA-SIC
$2s$ cross section come from? 

In order to answer this we need to take a close look at the single-electron
ground state LDA and LDA-SIC radial potentials. 
As pointed out earlier, in LDA formalism ``all" electrons of the system in 
the ground state experience the same potential defined by equation (1)
which for Na$_{20}$ is denoted by the dotted curve in figure 3. 
The potential, as is typical for a cluster, is flat in the interior region
(region of de-localized quasi-free electrons)
while showing a strong screening at the edge around $R_{\mbox{c}}$;  
the unphysical exponential decay
at the long range may be noted. Switching to the
LDA-SIC scheme, electrons in every orbital now feel a distinctly different potential (see
equation (2)) with an approximately correct long range behavior
as represented by four solid curves in figure 3.
In this group of four SIC potentials the ones for $1s$, $1p$,
and $1d$ look qualitatively similar to the LDA potential but 
are slightly deeper. The $2s$ potential, on the other hand, exhibits    
a unique feature: a strong local variation around the position $r=R_{\mbox{\scriptsize{n}}}$.
To pin down how this structure in the $2s$ LDA-SIC potential comes about we
need to focus on the SIC exchange correction $\vxcts$. This quantity, 
with reference to expression (3), can be explicitly written as:
\bea 
\vxcts &=& -\left(\frac{3\rho_{2s}(\vec{r}}{\pi}\right)^{1/3} -
        0.0333 \log \left[1+11.4\left(\frac{4\pi\rho_{2s}(\vec{r})}{3}\right)^{1/3}\right]
\eea
The $2s$ orbital density, $\rho_{2s}(\vec{r}) = |\phi_{2s}(\vec{r})|^2$, 
in the above equation,   
vanishes at $r=R_n$ as the $2s$ radial wavefunction passes through its node at $R_n$. 
Consequently, $\vxcts$ generates a cusp-like structure in the 
neighborhood of $R_n$
that shows up in the LDA-SIC potential profile for the $2s$ orbital. Since the behavior 
here is directly connected to the zero in the $2s$ electron density we stress that it must
also occur in any alternate prescription for $\vxcts$ different from formula (2).
We further emphasize
that this
structure is entirely an artifact of an externally imposed SIC in a purely local frame, 
which certainly is an approximation since a complete cancellation of self-interactions 
requires an appropriate non-local treatment of the electron-exchange 
phenomenon as in the HF formalism. In fact, a forced localization of the 
exchange (Fock) term in the HF scheme does indeed 
produce an infinite singularity in the potential at the zero of the corresponding 
one-electron state function [12].
Nevertheless, the structure from LDA-SIC has a direct bearing on the subsequent $2s$ photoionization 
matrix element by producing an unphysical oscillation. 

To behold the underlying mechanism let us consider the photoionization
dipole matrix element. We use for convenience the acceleration gauge representation of the dipole
interaction that involves the gradient of the potential seen
by the outgoing electron.  
After carrying out angular integration with the
assumptions of spherical
symmetry and unpolarized light, one is left with a reduced radial matrix
element for a dipole transition $nl \rightarrow \epsilon l'$ 
that in the acceleration formalism is $<\psi_{\epsilon l'}|dV/dr|\psi_{nl}>$.
Figure 4 shows that
the derivatives of both the LDA potential and the LDA-SIC $2s$ potential 
peak close to $r=R_c$.
In fact, the first derivative of any general cluster potential 
always peaks at the edge $R_c$, and therefore, the
overlap integral in the radial matrix element has dominating contribution 
coming from the edge [8]. Further,
for high enough energy $\psi_{\epsilon l'}$ can be described in the 
first Born picture as a spherical wave with asymptotic
form $\cos(k_{nl}r+\delta_{l'})$. This  
immediately suggests
that the matrix element will oscillate in the $k_{nl}$ space
 with roughly a frequency that is  
equal to the distance of the peak derivative point from the origin. 
 As a consequence, resulting
cross sections should exhibit 
an oscillation as a function of $k_{nl}$ with a 
frequence $2R_c$ (since the cross section is the squared modulus 
of the 
matrix element). As mentioned before, this effect is already known and can be related to the common
frequency peak in figure 2. However, something additional happens for the LDA-SIC case. The 
structure induced by the wavefunction node in the LDA-SIC potential for 
the $2s$ orbital produces a sharp discontinuity at $R_n$ in $dV_{2s}/dr$, as also seen in
figure 4.
Such a derivative-discontinuity induces a second oscillation in the      
respective overlap integral with a frequency about $R_n$ [13]. Subsequently, the $2s$
cross section with SIC acquires four oscillation frequencies: $R_c-R_n$, $2R_n$, $R_c+R_n$,
and $2R_c$ (see figure 2) as a result of    
the interference. Evidently, the first three frequencies are artificial 
being connected to the unphysical structure in the potential. Non-local  
HF, which
exactly eliminates electron self-interaction terms, is free from this effect (as is seen
from figures 1 and 2). 
Moreover, the qualitative agreement of HF with 
LDA suggests that SIC is practically unimportant at high enough photon-energy.
This is simply because with predominant contribution coming from the potential
edge for higher energies any improvement in the asymptotic behavior of the wavefunction
does not significantly influence the overlap integral. Therefore, for large systems
where HF becomes computationally impracticable
the usual LDA may be a safe choice in the high-energy regime. On the other hand, the fact that 
slower photoelectrons with their longer wavelength can hardly ``resolve" this nodal
structure explains why low-energy cross sections in LDA-SIC are practically
uncontaminated.   
It is 
also simple to understand that there is nothing special about $2s$ 
photoelectrons, for the effect must also be present in the case of   
subshells 
having more than one node.

The characteristic potential for any de-localized electron system, as in a metal cluster,
has a nearly flat interior region. Any rapid variation in this potential occurring
in a small range  
can, therefore, have considerable effect on the photoionization overlap integral by significantly 
altering the amplitude
of the continuum wave across this range. For atomic systems, however, electrons 
are far more localized owing to the strong nuclear attraction, and therefore,
wavefunctions are far more compact around the nucleus. The near-Coulombic
shape of a typical atomic potential with steep slope close to the origin can practically 
overwhelm any local variation as the one discussed in this paper. In order to verify this, we
applied LDA-SIC for some typical cases of 
atomic photoionization without any problem. 

It is true that SIC in the LDA frame induces certain extensiveness in the calculations
by making the potential state-dependent. One possible simplification is to average
over all such state-dependent potentials and use the averaged  
one for all electrons. We applied such an average-SIC
potential to examine whether or not the effect reduces. We find that 
not only the effect survives but that it also now substantially affects  photoelectrons 
from subshells without a node 
because the wavefunction overlap across the nodal zone is rather strong for them since their 
ground state wavefunctions are large in this region. However, it remains to be seen 
what happens if the potential is further approximated by a simplified-implementation
of SIC, namely, the optimized effective potential method [14]. 
Finally, it has
recently been found in the context of atoms that the independent particle model
breaks down for the high energy photoionization due to the interchannel coupling effect [15]. 
There is no {\em a\,priori} reason
to assume that this will not be the case for cluster systems, although no study has yet been made.
Nevertheless, in the future even if a multi-channel frame (namely, the time-dependent
LDA which is akin to the random-phase approximation) is needed to characterize the energetic
photoionization of clusters, this spurious effect will remain, at least qualitatively, 
and may also 
affect those channels whose single channel description is 
otherwise error-free. 

To summarize, we have shown  
that the theoretical analysis in the framework of 
LDA with SIC incorporated
may invoke unphysical strong qualitative variations in high energy photospectra 
of metal clusters for 
electrons emitted from a subshell with node(s); although there is no 
denying that LDA-SIC is one of the strong methodologies available  
to address low-energy processes. Through a comparison with the results
via non-local HF, that is intrinsically free from the self-interaction error, we conclude 
that the difficulty is connected to an inexact footing of SIC in the LDA formalism.  
Hence, it is important to choose appropriate theoretical techniques suitable for  
a given energy range
to avoid mis-interpretation of various effects in cluster photo-dynamical studies.

We thank Professor Steven T. Manson of GSU-Atlanta, USA, for making useful comments 
on the manuscript.

\newpage
\begin{figure}
\centerline{\psfig{figure=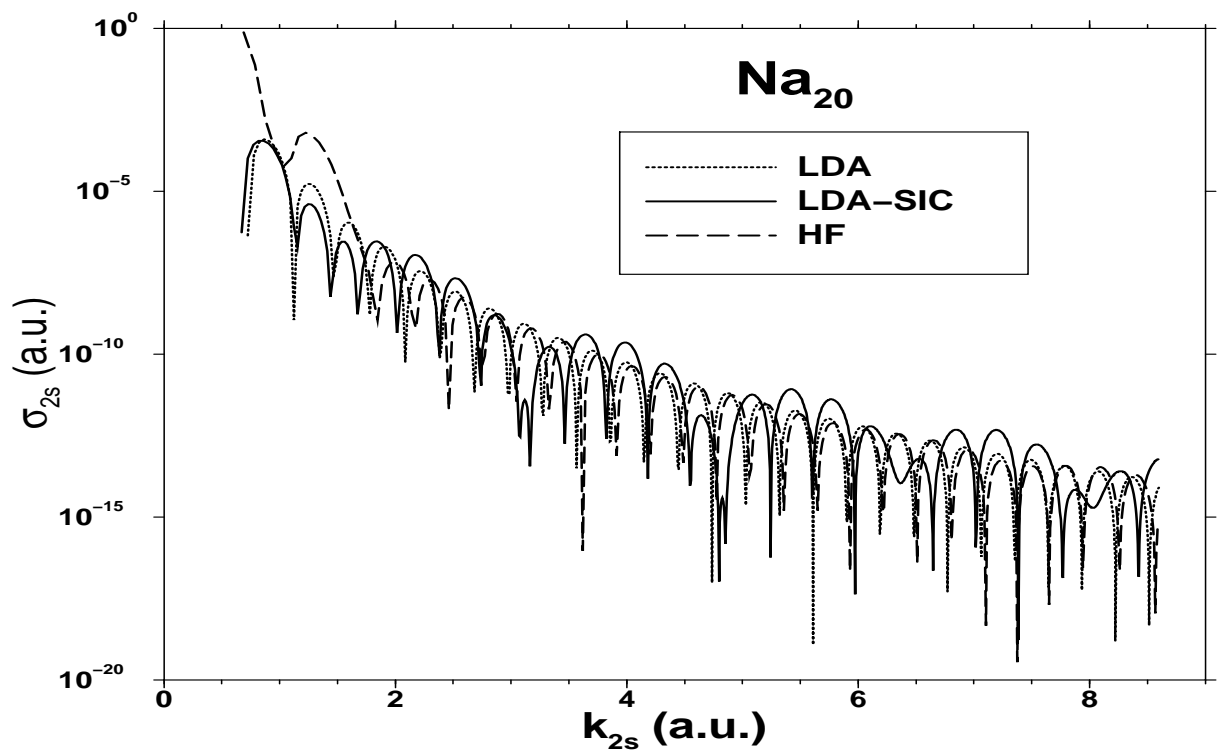,height=16cm,width=10cm,angle=-90}}
\caption{Photoionization cross sections for $2s$ subshell as a function of $2s$ 
photoelectron momentum calculated in LDA, LDA-SIC and HF approximations.}
\label{figure1}
\end{figure}
%\newpage
\begin{figure}
\centerline{\psfig{figure=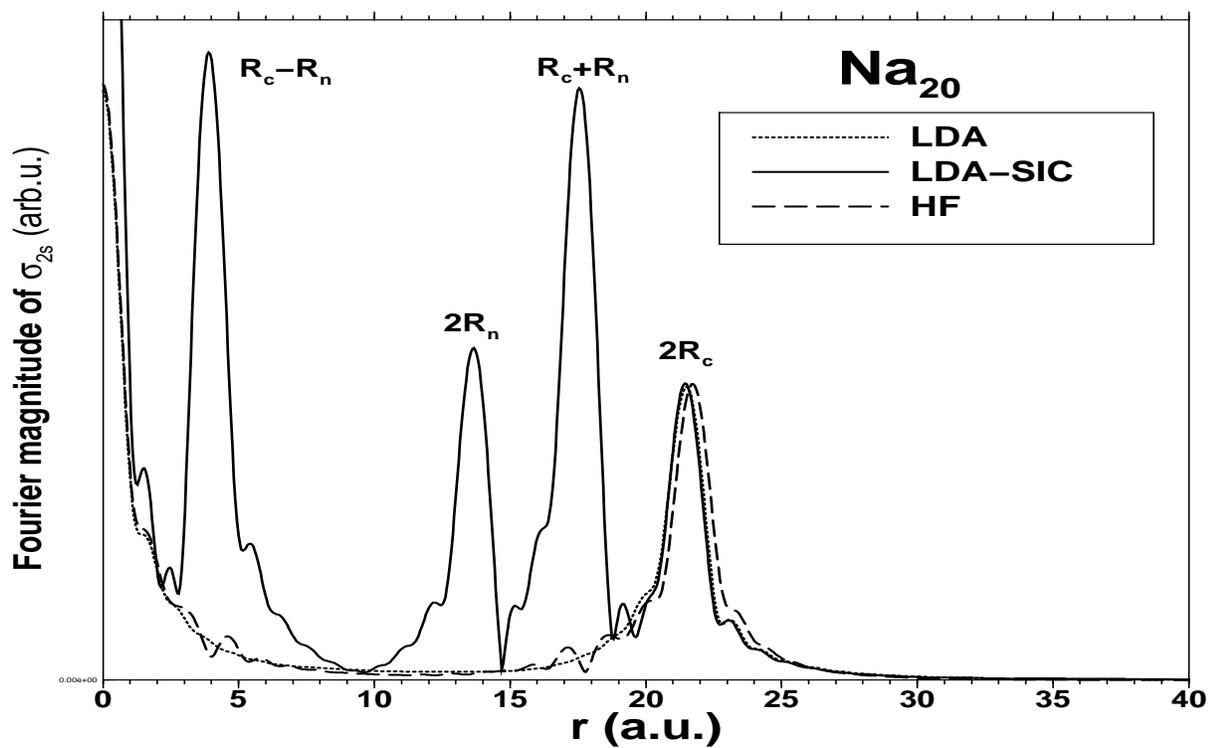,height=16.0cm,width=10cm,angle=-90}}
%\vskip 8cm
\caption{Fourier spectra of the same cross sections 
presented in figure 1.}
\label{figure2}
\end{figure}

\newpage
\begin{figure}[ht]
\centerline{\psfig{figure=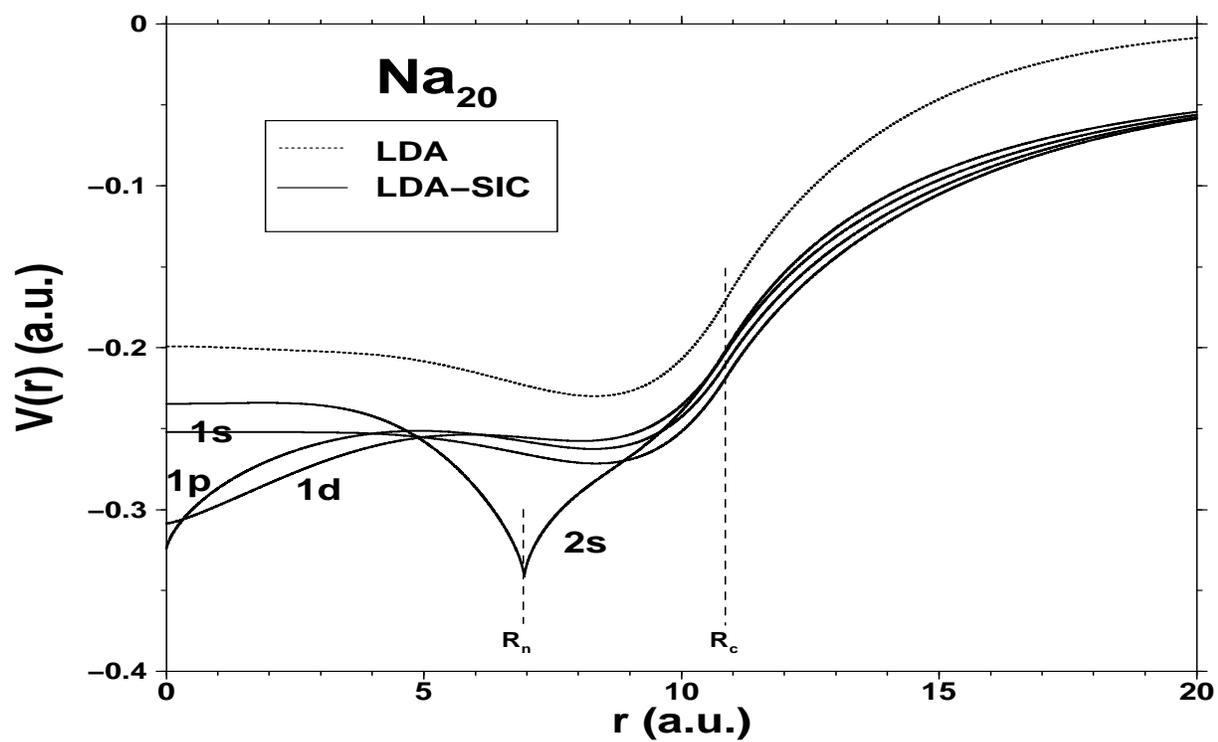,height=16cm,width=10cm,angle=-90}}
\caption{Comparison among LDA and four state-dependent 
LDA-SIC radial potentials. 
}
\label{figure3}
\end{figure}
%\newpage
\begin{figure}[h]
\centerline{\psfig{figure=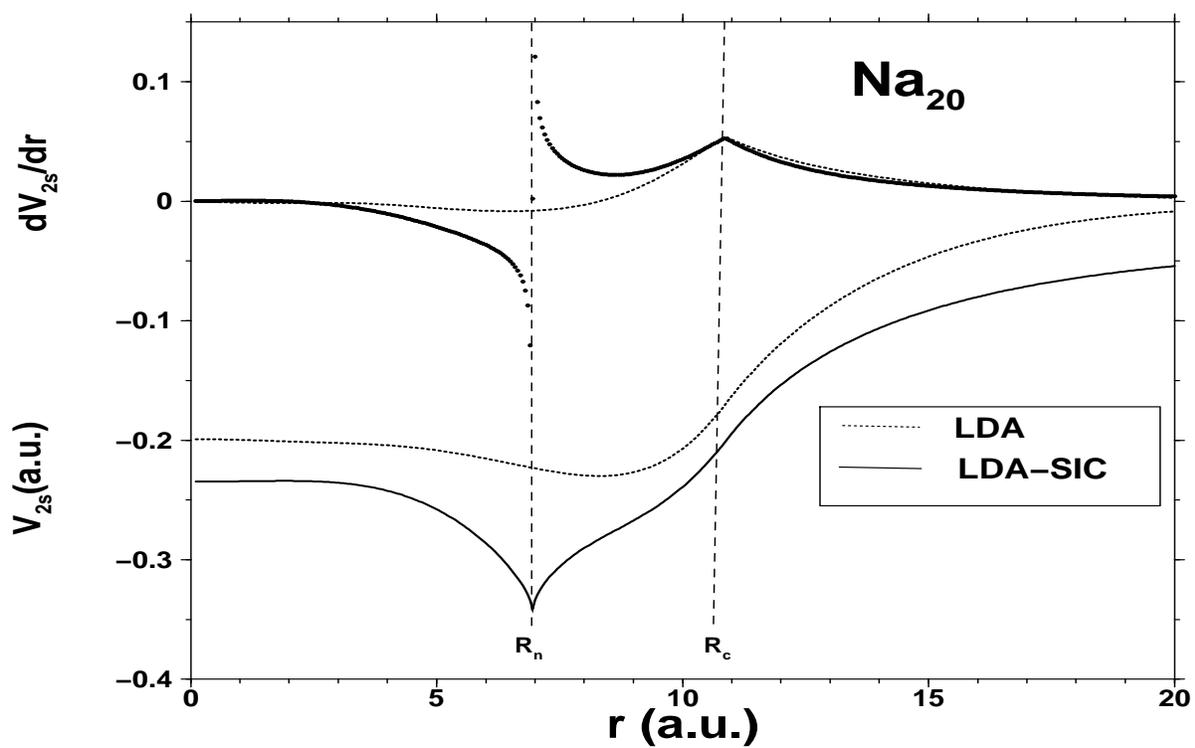,height=16cm,width=10cm,angle=-90}}
\caption{LDA and LDA-SIC-for-$2s$ potentials and their derivatives.
}
\label{figure4}
\end{figure}

\end{document}